\documentclass[aps,pra,superscriptaddress,floatfix,twocolumn,showpacs,10pt]{revtex4}
\usepackage[dvips]{graphicx}
\usepackage{amsmath}
\usepackage{times}

\begin{document}

\title{Purification of multipartite entanglement by local operations}

\author{Ming Yang\footnote{Corresponding Author}}
\email{ mingyang@ahu.edu.cn} \affiliation{School of Physics and
Material Science, Anhui University, Hefei 230039, People's Republic
of China}
\author{Fei Yan} \affiliation{School of Physics and Material
Science, Anhui University, Hefei 230039, People's Republic of China}
\author{Zhuo-Liang Cao}
\affiliation{Department of Physics {\&} Electronic Engineering,
Hefei Teachers College, Hefei 230061, People's Republic of China}
\affiliation{School of Physics and Material Science, Anhui
University, Hefei 230039, People's Republic of China}
\date{\today}
\begin{abstract}
Multipartite entanglement purification is revisited by using the
Local operations and classical communications(LOCCs). We demonstrate
our idea by considering the tripartite case, i.e. the purification
of tripartite entanglement. We express the general
tripartite entangled states in a special representation of total spin operators $J_{123}^{2}$ and $%
J_{12}^{2}$ of tripartite system with eigenvalues $15/4$ and $2$
respectively. This basis is a \textit{genuine basis} because it
consists of of all the genuine entangled states of tripartite
system. Our protocol is a recurrence one, and only two copies of the
initial mixed tripartite entangled states are needed in each round.
It is shown that if the initial fidelity is larger than a threshold
$0.4$, the purification process will succeed.The yield of the
current protocol is higher than the previous multipartite
entanglement purification protocols. As a by-product, we can get a
bipartite pure Bell state when the purification protocol fails for W
state. Our protocol also shows that there may be some special kind
of tripartite entanglement which belongs to neither W-type
entanglement nor GHZ-type entanglement.
\end{abstract}


\maketitle

\section{Introduction}

Entanglement is a indispensable resource for many quantum information
processing procedures, such as quantum teleportation\cite{teleportation},
quantum superdense coding \cite{densecoding}, quantum cryptography \cite%
{cryptography} and quantum computation\cite{computation} etc.
Especially, the multipartite entanglement plays an very important
role in quantum computation\cite{multipartitecompu1,
multipartitecompu2} and quantum
communincations\cite{multipartitecommu1, multipartitecommu2}. All of
the above procedures can be perfectly implemented with maximally
entangled bipartite or multipartite states only. But the
inevitability of the coupling between the entangled systems to its
environment will cause decoherence and disentanglement. Thus the
entangled states available in real situation are all mixed states,
which will decrease the efficiency and fidelity of quantum
communication and quantum computation. So how to enhance the
entanglement of the distributed mixed entangled states has been
becoming one of the highlighting research subjects in quantum
information field since the initial entanglement purification
protocol for bipartite mixed states\cite{purification}.

In the bipartite case, the basic purification idea is to extract
smaller number of entangled pairs from large number of initial
entangled pairs by Local operations and classical
communications(LOCCs), and the extracted
entangled pairs have more entanglement than the initial ones\cite%
{purification}. Purification of bipartite entanglement has been
realized in linear optical system\cite{pan1, pan2, gisin} and ionic
system\cite{romero, me, ionpuri}. Recently, the entanglement
purification protocols for higher dimensional entangled mixed states
have also been proposed\cite{higher1, higher2, higher3, higher4}.

In the multipartite case, the situation is a little bit different.
Multipartite system can be entangled in different ways, i.e. there
exist many different kinds of genuine entanglements, and all states
of each kind are considered to be equivalent under invertible local
operations. In the three qubits case, there are two kinds of genuine
tripartite entanglement, the Greenberger-Horne-Zeilinger(GHZ)
entanglement and the W entanglement\cite{wstate1}. The local
conversion between GHZ state and W state is only possible in an
approximate way\cite{wstate2}. W. D\"{u}r \textit{et al} generalized
the bipartite entanglement purification to the multipartite case,
such as purification protocols for generalized GHZ state, cluster
state\cite{multipartitecompu2} and various quantum error-correcting
code, all of which belong to the class of graph
states\cite{graphstate1, graphstate2, graphstate3}. They also showed
that the direct multipartite entanglement purification protocols are
more efficient than the approaches based on bipartite entanglement
purification. But these protocols does not apply to the mixed W
state. The reasons are twofold. First, the
entanglement of W state is different from the other types of entanglements%
\cite{wstate1}. Secondly, the expression of a general mixed W state is not
unique due to the existence of different kinds of genuine multipartite
entanglement. Akimasa Miyake and Hans J. Briegel constructed a W basis with
all of its states being equivalent to each other under local unitary
transformations, and the noisy W states resulting from typical decoherence
were distilled by complementary stabilizer measurements rather than the
standard CNOT operations and projection measurements used in the bipartite
entanglement purification protocol\cite{wpuri}. In this paper, we will
present an alternative construction of the expansion basis for a general
mixed W state, which consists of the two kinds of genuine tripartite
entangled states, GHZ states and W states. So we call this basis \textit{%
genuine basis}. We generalize the standard bipartite entanglement
purification protocol to the general mixed W state expressed in terms of the \textit{%
genuine basis.} Our protocol is a recurrence one, and only two
copies of the original mixed state are needed in each round. Every
user will have two particles from the two different entangled
triples undergo a CNOT operation. Then a projection measurement will
be carried out on the target particle by each user. Through
classical communication, all the users will compare their
measurement results and decide whether keep the left particles or
not. If the measurement results of the three users are same, the
entanglement of the left triple is enhanced i.e. the fidelity with
respect to the standard W states can be amplified via LOCCs. If the
measurement results are not same, the purification protocol for W
state fails. But for some measurement results, two of the three
source particles will be left in a pure bipartite Bell state. This
will decrease the entanglement waste of the ordinary multipartite
purification protocols. This paper is organized as follows: in
section II, we will give the detailed \textit{genuine basis} of a
tripartite
system and the general mixed W state expressed in terms of the \textit{%
genuine basis}; The purification process for this general mixed W
state will be discussed in section III. Some special cases about the
mixed W state will be discussed in section IV; Section V is the
conclusion section.

\section{\textit{Genuine Basis}}

Three qubits can be entangled in two inequivalent ways, and GHZ states and W
states are remarkable representatives of them\cite{wstate1}. W state is more
robust than GHZ state against the loss of one qubit. GHZ states and W states
are the only two kinds of tripartite genuine entangled states. From these
two genuine entangled states we construct the following \textit{genuine basis%
} for tripartite systems:
\begin{subequations}
\begin{equation}
\left\vert GB^{1-3}\right\rangle =\left[ \left\vert 001\right\rangle
+\omega^{n}_{3} \left\vert 010\right\rangle +\omega^{2n}_{3}
\left\vert 100\right\rangle \right]/\sqrt{3} , \label{GB1-3}
\end{equation}
\begin{equation}
\left\vert GB^{4-6}\right\rangle =\left[ \left\vert 110\right\rangle
+\omega^{n}_{3}\left\vert 101\right\rangle +\omega^{2n}_{3}
\left\vert 011\right\rangle\right]/\sqrt{3} , \label{GB4-6}
\end{equation}
\begin{equation}
\left\vert GB^{7}\right\rangle =\left( \left\vert 000\right\rangle
+\left\vert 111\right\rangle \right)/\sqrt{2} ,  \label{GB7}
\end{equation}
\begin{equation}
\left\vert GB^{8}\right\rangle =\left( \left\vert 000\right\rangle
-\left\vert 111\right\rangle \right)/\sqrt{2} ,  \label{GB8}
\end{equation}
\end{subequations}
where $n=0,1,2$, $\left\vert GB^{1-3}\right\rangle $ denote the \textit{%
genuine basis }$1,2,3$ corresponding to $n=0,1,2$, and the same
convention applies to the $\left\vert GB^{4-6}\right\rangle $.
$\omega_{3}=\exp\left(-i2\pi/3\right)$.The $\left\vert
GB^{7}\right\rangle $ and $\left\vert GB^{8}\right\rangle $ are two
GHZ states. One can easily check that this set of states constructs
an orthogonal and complete set of basis for tripartite systems. The
physical meaning of the above eight states can be understood in the
following way. The first set of three states in Eq.(\ref{GB1-3}) are
from the standard W state with different phases, and they are
orthogonal with each other; The second set of three states in
Eq.(\ref{GB4-6}) are from the flipped W state with different phases,
and they are also orthogonal with each other; The last two states in
Eqs.(\ref{GB7},\ref{GB8}) are two standard GHZ states from two
product states $\left\vert 000\right\rangle $ and $\left\vert
111\right\rangle $ with different phases. So the four basic states
of the \textit{genuine basis} are $\frac{1}{\sqrt{3}}\left(
\left\vert 001\right\rangle +\left\vert
010\right\rangle +\left\vert 100\right\rangle \right) $, $\frac{1}{\sqrt{3}}%
\left( \left\vert 110\right\rangle +\left\vert 101\right\rangle
+\left\vert 011\right\rangle \right) $, $\left\vert 000\right\rangle
$ and $\left\vert 111\right\rangle $, which are the four
simultaneous eigenvectors of total spin operators $J_{123}^{2}$ and
$J_{12}^{2}$ of tripartite systems with eigenvalues $15/4$ and $2$
respectively.

Using the \textit{genuine basis} we can write the general tripartite
mixed entangled state in the following form:
\begin{equation}
\rho =\sum^{8}_{i=1}C_{i}\left\vert GB^{i}\right\rangle \left\langle
GB^{i}\right\vert, \label{generalW}
\end{equation}%
which can describe the state of a distributed entangled quantum
ensemble and $C_{i}(i=1,2,\cdots ,8)$ are real numbers satisfying
normalization condition. For the sake of simplicity, we suppose
$C_{1}$ is the biggest coefficient. One can find that the first two
states $\frac{1}{\sqrt{3}}\left( \left\vert 001\right\rangle
+\left\vert 010\right\rangle +\left\vert 100\right\rangle \right) $ and $%
\frac{1}{\sqrt{3}}\left( \left\vert 110\right\rangle +\left\vert
101\right\rangle +\left\vert 011\right\rangle \right) $ of the four
basic states of \textit{genuine basis} are standard W states, so we
call this general mixed state as mixed W state. Because W state has
more robust entanglement than GHZ state, we are interested in
extracting W state from this mixed state. This is another reason why
we suppose $C_{1}$ is the biggest coefficient. For now, we use
\textit{W fraction} of the mixed state to indicate the entanglement
of it, i.e. $F_{W}=$max$\left\langle W\right\vert \rho \left\vert
W\right\rangle $\textit{\ }with the maximum being taken over the
first six state in the \textit{genuine basis}. So the
\textit{W fraction} of the general mixed W state in Eq.(\ref{generalW}) is $%
F_{W}=C_{1}$.

\section{Purification protocol for the general mixed W states}

\bigskip Because the local operation and classical communication can not
creat remote entanglement, the idea of purifying this mixed state is
to transfer entanglement from big ensembles to small ones with only
LOCCs. Next, we will discuss the purification process in detail.

Firstly, three remote users Alice, Bob and Charlie possess a distributed
entangled quantum ensemble whose state can be described by Eq.(\ref%
{generalW}). To get a state with increased \textit{W fraction}, the
three users must operate on two triples of particles of this
distributed ensemble:
\begin{subequations}
\begin{equation}
\rho _{ABC}^{s}=\sum^{8}_{i=1}C_{i}\left\vert GB^{i}\right\rangle
^{s}\left\langle GB^{i}\right\vert, \label{rhos}
\end{equation}
\begin{equation}
\rho _{ABC}^{t}=\sum^{8}_{i=1}C_{i}\left\vert GB^{i}\right\rangle
^{t}\left\langle GB^{i}\right\vert, \label{rhot}
\end{equation}%
\end{subequations}
where the subscripts $A,B,C$ denote the three users Alice, Bob and
Charlie, respectively, and the superscripts $s,t$ denote the source
and target triples, respectively. Each user will carry out a
controlled not(CNOT) operation on the two particles from two mixed
states, i.e. the trilateral XOR operation(TXOR) which is analogous
to the bilateral XOR operation(BXOR) in the bipartite entanglement
purification protocol\cite{purification}. After the TXOR operation
the target particles will be measured, and the three users will
compare their measurement results through classical communication
and decide whether keep the source triple or not. To get the
explicit expression of the state of the two triples after TXOR
operation, we can treat the initial state of the two triples as the
probabilistic mixture of $64$ pure states: $\left\vert
GB^{i}\right\rangle^{s} \left\vert GB^{j}\right\rangle^{t} $ with a
probability $C_{i}C_{j}$, $i,j=1,2,\cdots,8$.

Straightforward calculation shows that if the measurement results of
the three users are same, the left triple are still in a tripartite
entangled mixed state, which can have a higher \textit{W fraction
}provided some restrictions on the initial \textit{W fraction}. That
is to say the purification protocol can succeed with some
probability. For measurement $\left\vert 000\right\rangle
_{ABC}^{t}$, we can get a new mixed W state with some probability
and new \textit{W\ fraction}:
\begin{widetext}
\begin{subequations}
\begin{equation}
P_{000}=\left[2\left( C_{1}+C_{2}+C_{3}\right) ^{2}+2\left(
C_{4}+C_{5}+C_{6}\right) ^{2}+3\left( C_{7}+C_{8}\right)
^{2}\right]/6, \label{prob000}
\end{equation}
\begin{equation}
F_{W}^{000}=\frac{2C_{1}^{2}+4C_{2}C_{3}}{2\left(
C_{1}+C_{2}+C_{3}\right) ^{2}+2\left( C_{4}+C_{5}+C_{6}\right)
^{2}+3\left( C_{7}+C_{8}\right) ^{2}}.  \label{fidelity000}
\end{equation}
\end{subequations}
\end{widetext}

\begin{figure}[tbp]
\includegraphics[width=0.5\textwidth]{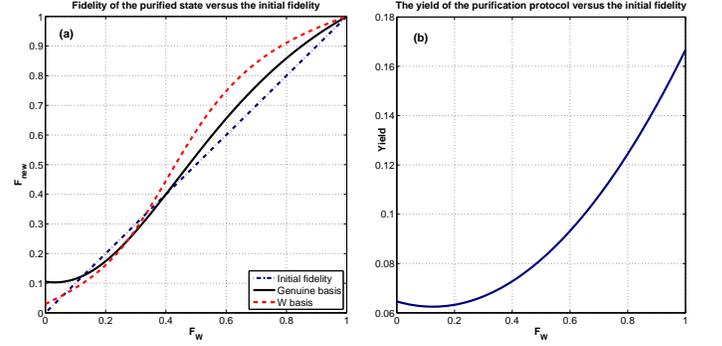}
\caption{\label{Fnew}(a)The \emph{W fraction} of the purified state
in the current protocol(solid line) and the fidelity of the purified
state in Ref.\cite{wpuri}(dashed line)varying with the fidelity of
the initial state(dash-dotted line); (b)The yield of the current
protocol, i.e. the ratio of the number of the surviving copies to
that of the used copies.}
\end{figure}

\bigskip To simplify the calculation, without loss of generality, we can let $%
C_{i}=\frac{1-C_{1}}{7}, i = 2,3,\cdots,8$ by a \textit{%
random trilateral rotation} on the initial mixed state in Eq.(\ref{generalW}%
), which is analogous to the \textit{random bilateral rotation} of
the bipartite entanglement purification protocol\cite{purification}.
Under this situation, the initial mixed W state takes a concise
form:
\begin{equation}
\rho _{cs}=C_{1}\left\vert GB^{1}\right\rangle \left\langle
GB^{1}\right\vert +\frac{1-C_{1}}{7}\left( I-\left\vert GB^{1}\right\rangle
\left\langle GB^{1}\right\vert \right) ,  \label{conciseW}
\end{equation}%
Now the \textit{W\ fraction} in Eq.(\ref{fidelity000}) becomes: $F_{W}^{000}=%
\frac{51C_{1}^{2}-4C_{1}+2}{40C_{1}^{2}-10C_{1}+19}$. Obviously, if $%
C_{1}>2/5$, $F_{W}^{000}-F_{W}>0$, which means an amplification of
the entanglement of the initial mixed W state. The new \textit{W\
fraction }$F_{W}^{000}$ and the yield($P_{000}/2$) varying with the
initial \textit{W\ fraction }$F_{W}$ are shown in Fig.(\ref{Fnew}a)
and Fig.(\ref{Fnew}b) respectively. The Fig.(\ref{Fnew}a) shows that
the fidelity of the purified mixed W state in Ref.(\cite{wpuri}) is
higher than ours new\textit{\ W fraction}, but the yield of the
current protocol is larger than that of Ref.\cite{wpuri}. The high
yield means a great reduction of the number of iterations before the
\textit{W fraction} approaching one.

For measurement result $\left\vert 111\right\rangle _{ABC}^{t}$,
although we can get a tripartite entangled mixed state with new \textit{W\ fraction }$%
F_{W}^{111}$, the new \textit{W\ fraction }$F_{W}^{111}$ never
exceeds
the initial \textit{W\ fraction }$F_{W}$ because of the assumption that $%
C_{1}$ is the biggest coefficient. So this result does not
contribute to the purification of general mixed W states.

For other six measurement results, the state of the left triple only
has bipartite entanglement, which indicates the fail of the
purification for mixed tripartite W states.

\section{Some discussions on the results of the purification protocol}

Although the tripartite purification will fail for the above
mentioned six unwanted measurement results, the left triple of
particles will still have bipartite entanglement in these six cases.
In general, this bipartite entanglement is a mixed entanglement.
From the detailed results of the purification protocol in section
III, we can see that if the initial state satisfies some conditions,
this left bipartite entanglement can be pure Bell entanglement
without measurement on the third one. For example, if the initial
state is a mixture of two states $\left\vert GB^{1}\right\rangle $,
$\left\vert GB^{4}\right\rangle ($or $\left\vert GB^{2}\right\rangle
$, $\left\vert GB^{5}\right\rangle $ or $\left\vert
GB^{3}\right\rangle $, $\left\vert GB^{6}\right\rangle )$ , only
$C_{1}$, $C_{4}($or $C_{2}$, $C_{5}$ or $C_{3}$, $C_{6})$ are not
zero, and all the other coefficients are vanishing. Then, for the
measurement results $\left\vert 100\right\rangle _{ABC}^{t}$,
$\left\vert 010\right\rangle _{ABC}^{t}$, $\left\vert
001\right\rangle _{ABC}^{t}$ , the two particles of the left triple
can be in a pure Bell state with the third particle being factorized
out naturally. Another example is a very
interesting one where we suppose $C_{1}=C_{4}=1/2($or $C_{2}=C_{5}$, or $%
C_{3}=C_{6})$ and the other coefficients are vanishing. In these
cases, the initial states are all in a total mixture of two
orthogonal basis states from the first six states of the
\textit{genuine basis}(For example, $\rho _{mix}=\frac{1}{2}\left(
\left\vert GB^{1}\right\rangle \left\langle GB^{1}\right\vert
+\left\vert GB^{4}\right\rangle \left\langle GB^{4}\right\vert
\right) $). Intuitively, there should be no entanglement in these
total mixtures. Counterintuitively, if we apply the above
purification protocol on these two mixtures, we will get pure
bipartite Bell state for the measurement results on the targets
$\left\vert 110\right\rangle _{ABC}^{t}$, $\left\vert 011\right\rangle _{ABC}^{t}$, $%
\left\vert 101\right\rangle _{ABC}^{t}$. This point can be
understood from analyzing the entanglement property of the total
mixture $\rho _{mix}$. Although, up to now, we do not have a very
rigorous definition of the entanglement measure of multipartite
system, we can detect whether there is some kind of entanglement or
not by using entanglement witness\cite{witness1, witness2}. This is
done by constructing a witness operator that can detect whether the
state has the desired genuine entanglement $\left\vert
\psi \right\rangle $, and this witness operator takes the form $\mathcal{W}%
_{\psi }=\alpha I-\left\vert \psi \right\rangle \left\langle \psi
\right\vert $, with $\alpha =\underset{\left\vert \phi \right\rangle \in B}{%
\max }\left\vert \left\langle \phi \right. \left\vert \psi \right\rangle
\right\vert ^{2}$. Here $B$ denotes the set of biseparable states. If the
expectation value of the witness operator is negative in some state, we can
say this state has multipartite entanglement of the kind of $\left\vert \psi
\right\rangle $, otherwise these is no genuine entanglement in this state.
For example, ones used to use the witness operator $\mathcal{W}_{W}=\frac{1}{%
2}I-\left\vert GHZ\right\rangle \left\langle GHZ\right\vert $ for
tripartite W state, and $\mathcal{W}_{GHZ}=\frac{3}{4}I-\left\vert
GHZ\right\rangle \left\langle GHZ\right\vert $ for tripartite GHZ
state\cite{witness1, witness2}. If we apply these two tripartite
entanglement witness operators on the above mentioned total mixture
states $\rho _{mix}$, we will find that these total mixture $\rho
_{mix}$ do not carry any genuine tripartite entanglement. But, if we
trace out one of the three particles, bipartite entanglement is
there between the two left particles. So, it is possible to
extract a pure Bell state from the above mentioned two total mixtures states $%
\rho _{mix}$.

\bigskip Using the witness operator we can detect the entanglement of the
initial general mixed W state in Eq.(\ref{conciseW}). Direct calculations
show that $Tr\left( \mathcal{W}_{W}\rho _{cs}\right) <0$ is found when $%
C_{1}>\frac{13}{20}$. That is to say, only when the \textit{W\
fraction }is over $\frac{13}{20}$ the state in Eq.(\ref{conciseW})
posses W-type entanglement, which coincides with the result of
Ref.\cite{witness1}. But the threshold of our purification protocol
is $\frac{2}{5}$, which is smaller than $\frac{13}{20}$. The fact is
that, by using our purification protocol, one always can get a mixed
W state with \textit{W\ fraction } over $\frac{13}{20}$, which means
that we distill W-type entanglement from non-W-type entanglement. In
this sense, we say that the initial state with \textit{W\ fraction
}being in the range $\frac{2}{5}<F_{W}<\frac{13}{20}$ has some
special kind of tripartite entanglement which belongs to neither
W-type entanglement nor GHZ-type entanglement($Tr\left( \mathcal{W}%
_{GHZ}\rho _{cs}\right) >0$ for $\frac{2}{5}<C_{1}<1$). At the same time,
our threshold is also different from that($\frac{1}{3}$) of Ref.\cite{wpuri}%
, and this point may caused by the reason that we used a type of
initial entangled mixed W state which is different from that of
Ref.\cite{wpuri}. This may be another evidence of the existence of
the above mentioned special tripartite entanglement.

Generalization of the current tripartite purification protocol to
the multipartite case is not straightforward, because it is not easy
to find the \textit{genuine basis} for the multipartite case.
Anyway, we found the \textit{genuine basis }for the four-particle
system:
\begin{subequations}
\begin{eqnarray}
\left\vert GB_{4}^{1}\right\rangle &=&\frac{1}{2}\left( \left\vert
0001\right\rangle +\left\vert 0010\right\rangle +\left\vert
0100\right\rangle +\left\vert 1000\right\rangle \right), \\
\left\vert GB_{4}^{2}\right\rangle &=&\frac{1}{2}\left( \left\vert
0001\right\rangle +\left\vert 0010\right\rangle -\left\vert
0100\right\rangle -\left\vert 1000\right\rangle \right), \\
\left\vert GB_{4}^{3}\right\rangle &=&\frac{1}{2}\left( \left\vert
0001\right\rangle -\left\vert 0010\right\rangle +\left\vert
0100\right\rangle -\left\vert 1000\right\rangle \right), \\
\left\vert GB_{4}^{4}\right\rangle &=&\frac{1}{2}\left( \left\vert
0001\right\rangle -\left\vert 0010\right\rangle -\left\vert
0100\right\rangle +\left\vert 1000\right\rangle \right),\\
\left\vert GB_{4}^{5}\right\rangle &=&\frac{1}{2}\left( \left\vert
1110\right\rangle +\left\vert 1101\right\rangle +\left\vert
1011\right\rangle +\left\vert 0111\right\rangle \right), \\
\left\vert GB_{4}^{6}\right\rangle &=&\frac{1}{2}\left( \left\vert
1110\right\rangle +\left\vert 1101\right\rangle -\left\vert
1011\right\rangle -\left\vert 0111\right\rangle \right), \\
\left\vert GB_{4}^{7}\right\rangle &=&\frac{1}{2}\left( \left\vert
1110\right\rangle -\left\vert 1101\right\rangle +\left\vert
1011\right\rangle -\left\vert 0111\right\rangle \right), \\
\left\vert GB_{4}^{8}\right\rangle &=&\frac{1}{2}\left( \left\vert
1110\right\rangle -\left\vert 1101\right\rangle -\left\vert
1011\right\rangle +\left\vert 0111\right\rangle \right),\\
\left\vert GB_{4}^{15}\right\rangle &=&\left( \left\vert
0000\right\rangle +\left\vert 1111\right\rangle \right)/\sqrt{2}, \\
\left\vert GB_{4}^{16}\right\rangle &=&\left( \left\vert
0000\right\rangle -\left\vert 1111\right\rangle \right)/\sqrt{2},
\end{eqnarray}
\begin{widetext}
\begin{eqnarray}
\left\vert GB_{4}^{9}\right\rangle &=&\left( \left\vert
1100\right\rangle +\left\vert 1010\right\rangle +\left\vert
1001\right\rangle +\left\vert 0110\right\rangle +\left\vert
0101\right\rangle +\left\vert 0011\right\rangle \right)/\sqrt{6}, \\
\left\vert GB_{4}^{10}\right\rangle &=&\left( \left\vert
1100\right\rangle +\omega ^{10}\left\vert 1010\right\rangle +\omega
^{8}\left\vert 1001\right\rangle -\left\vert 0110\right\rangle
+\omega ^{4}\left\vert 0101\right\rangle +\omega ^{2}\left\vert
0011\right\rangle
\right)/\sqrt{6}, \\
\left\vert GB_{4}^{11}\right\rangle &=&\left( \left\vert
1100\right\rangle +\omega ^{8}\left\vert 1010\right\rangle +\omega
^{4}\left\vert 1001\right\rangle +\left\vert 0110\right\rangle
+\omega ^{8}\left\vert 0101\right\rangle +\omega ^{4}\left\vert
0011\right\rangle
\right)/\sqrt{6}, \\
\left\vert GB_{4}^{12}\right\rangle &=&\left( \left\vert
1100\right\rangle -\left\vert 1010\right\rangle +\left\vert
1001\right\rangle -\left\vert 0110\right\rangle +\left\vert
0101\right\rangle -\left\vert 0011\right\rangle \right)/\sqrt{6}, \\
\left\vert GB_{4}^{13}\right\rangle &=&\left( \left\vert
1100\right\rangle +\omega ^{4}\left\vert 1010\right\rangle +\omega
^{8}\left\vert 1001\right\rangle +\left\vert 0110\right\rangle
+\omega ^{4}\left\vert 0101\right\rangle +\omega ^{8}\left\vert
0011\right\rangle
\right)/\sqrt{6}, \\
\left\vert GB_{4}^{14}\right\rangle &=&\left( \left\vert
1100\right\rangle +\omega ^{2}\left\vert 1010\right\rangle +\omega
^{4}\left\vert 1001\right\rangle -\left\vert 0110\right\rangle
+\omega ^{8}\left\vert 0101\right\rangle +\omega ^{10}\left\vert
0011\right\rangle \right)/\sqrt{6},
\end{eqnarray}
\end{widetext}
\end{subequations}
where $\omega =\exp \left( \frac{i\pi }{6}\right) $. One can easily
design the purification protocol for the four-partite mixed
entangled states along the lines we suggested here.

\section{Conlusions}

We construct a complete basis from all the two kinds tripartite
genuine entangled states, and we call it \textit{genuine basis}. The
general mixed W state of tripartite system is expressed in terms of
this \textit{genuine basis}. The \textit{W fraction} can be enhanced
by LOCCs, i.e. it can be purified. Multipartite entanglement witness
shows that there exists a special kind of tripartite entanglement
which belongs to neither W-type entanglement nor GHZ-type
entanglement, but we can extract W-type entanglement from this kind
of entangled states. The yield of the protocol is higher than the
previous multipartite entanglement purification protocol for mixed W
states. As a by-product, we can get bipartite pure Bell states when
the purification protocols fails for tripartite entangled states,
depending on the initial mixed states.

\begin{acknowledgements}
\bigskip This work is supported by National Natural Science Foundation of
China (NSFC) under Grant Nos: 60678022 and 10704001, the Specialized
Research Fund for the Doctoral Program of Higher Education under Grant No.
20060357008, Anhui Provincial Natural Science Foundation under Grant No.
070412060, the Talent Foundation of Anhui University, and Anhui Key
Laboratory of Information Materials and Devices (Anhui University).
\end{acknowledgements}

\end{document}